# Thermodynamic Approach for Deciphering Magneto-Structural Phase Transitions: Proof of Concept in Heusler Alloys

Eleonora Rusconi[a], Lorenzo Gallo[a], Victor A. L'vov[b], Anna Kosogor[b,c,d], Simone Fabbrici[a], Giovanna Trevisi[a], Francesco Cugini[a,e], Massimo Solzi[a,e], Thomas Schrefl[d], Franca Albertini[a]

[a] Institute of Materials for Magnetism and Electronics, National Research Council, Parco Area delle Scienze 37/A, 43124 Parma, Italy
[b] V.G. Baryakhtar Institute of Magnetism of the NAS of Ukraine, Ukraine
[c] University of Vienna, Faculty of Physics, Austria
[d] University for Continuing Education Krems, Austria
[e] University of Parma, Department of Mathematical, Physical and Computer Sciences, Parco Area delle Scienze 7/A, 43124 Parma, Italy

**ABSTRACT**

Ferromagnetic solids acquire nontrivial magnetic and caloric properties when the temperature of the structural phase transition approaches the Curie point. Deciphering magneto-structural transitions, i.e. determining their characteristic temperatures and elucidating the related properties, remains challenging. In the present paper, three types of transformational behaviour of $Ni_{50}Mn_{25-x}Cu_xGa_{25}$ ($x$ = 6.25, 6.5, 6.75, 7) and $Ni_{50.5}Mn_{18.5}Cu_{6.5}Ga_{24.5}$ alloys have been identified, arising from small variations in chemical composition: (i) structural martensitic transformation (MT) in the ferromagnetic phase; (ii) magneto-structural phase transition from paramagnetic austenite to ferromagnetic martensite; (iii) MT in paramagnetic phase. The temperature-dependent values of magnetization, $M(T)$, and of magnetic susceptibility, $\chi(T)$, were measured for each alloy. A novel thermodynamic analysis was used to determine the Curie points and MT temperatures. The novelty lies in considering the interplay between structural and magnetic characteristics of the alloys through the impact of the structural transition on the spin-exchange parameter. The theoretical analysis of experimental data revealed that this impact results in a large difference ($\geq$ 50 K) between the Curie temperatures computed for the austenitic and martensitic states of each alloy. The characteristic temperatures, corresponding to the extrema of the $dM(T)/dT$ and $\chi(T)$ functions, were calculated. The correlation of these temperatures with the Curie temperatures and the MT temperatures is not straightforward and depends strongly on the type of transformational behaviour (i) – (iii). The proposed approach provides a robust framework for extracting unmeasurable characteristic temperatures from standard magnetization data, applicable to ferromagnetic Heusler systems and other multiferroic ferromagnetic materials.

**HIGHLIGHTS**

- Cu doping of Ni-Mn-Ga adjusts the martensitic transition temperatures to Curie point
- The structural and magnetic characteristics of Ni-Mn-Cu-Ga are strongly intertwined
- Standard methods for determining the Curie temperature are not applicable
- An adequate method for determining the characteristic temperatures is presented
- The presented method is applicable to different multiferroic materials

## 1. INTRODUCTION

Multifunctional materials undergo a variety of structural phase transitions (see e.g., [1] and references therein), which, when intertwined with magnetic degrees of freedom, make it difficult to decipher the underlying physics. Determining their magnetic properties, i.e., Curie and structural transition temperatures, requires accounting for different characteristics and how they interrelate, such as their chemical composition, and the degree of chemical ordering, symmetry of the crystal lattice and microstructure, magnetoelastic coupling, the peculiarities of electron and phonon spectra. Moreover, these characteristics change in response to external stimuli, e.g., magnetic field, axial stress, or hydrostatic pressure, further complicating the picture.

Metamagnetic Heusler alloys are a prime example of multifunctional materials. They belong to the family of multiferroic materials undergoing ferroelastic phase transitions, such as Fe-Rh [2, 3], La-Fe-Si [4, 5], $(Mn,Fe)_2(P,Si)$-type [6 ,7, 8] compounds, but are unique in combining shape memory properties, magnetocaloric and elastocaloric effects [9-12], large magnetoresistance [13,14] and giant magnetic-field-induced strain (MFIS) [15-17].

The quasi-stoichiometric Ni-Mn-Ga Heusler alloys are the most intensively studied ferromagnetic shape memory alloys (FSMA) due to the observation of large (exceeding 10%) MFIS [15-22], and giant magnetocaloric effect [9, 22, 23], which make them promising for a variety of applications from magnetic refrigeration to actuation, sensing and biomedical purposes [24, 25, 26]. The multifunctionality of magnetic shape memory Heusler alloys is strongly related to their structural phase transformation from a high-temperature high-symmetry phase (austenite) to a low-temperature low-symmetry phase (martensite), or to magneto-structural phase transitions, where the material transforms between these two crystalline phases while also undergoing a change in magnetic ordering. Phase transition temperatures can be tuned by changing the chemical composition of alloys and by appropriate thermal treatments, to obtain suitable alloys for different applications.

The present paper focuses on the characterization of ferromagnetic alloys exhibiting a first-order ferroelastic phase transition from the high-temperature austenitic to the low-temperature martensitic phase. The martensitic transformation (MT) temperature, the Curie temperature, the temperature dependence of magnetization $M(T)$, and magnetic susceptibility $\chi(T)$, are among the most important characteristics of these materials. The problems arising in the determination of MT temperatures are discussed in Ref. [27] and references therein. By contrast, a fundamental difficulty in determining the Curie temperature arises when it is close to the MT temperature. As it is well known, the Curie temperature corresponds to the onset of magnetic ordering upon cooling of a paramagnetic (PM) solid and is often determined from the minimum point $(T_{\min}, M_{\min})$ of the temperature derivative of the magnetization ($dM / dT$) (see e.g. Ref. [28]). However, this method is not valid if the Curie temperature is close to the MT temperature [29]. Two characteristic temperatures, the Curie temperature of austenite and the Curie temperature of martensite, must be considered to explain the temperature dependences of magnetization measured for different Ni-Mn-Ga alloys, as introduced in Ref. [30]. It was emphasized that the temperature dependence of magnetization is strongly conditioned by the sequence of phase transitions, which can be inherent to each specific alloy. Specifically, for the phase transition from paramagnetic austenite to ferromagnetic (FM) martensite, the Curie temperature coincides with the MT temperature [30].

The MT temperature can be purposefully changed by adjusting the chemical composition of the Ni-Mn-Ga-based alloys. Albertini *et al*. reported the possibility of obtaining a direct magneto-structural PM austenite – FM martensite phase transition over a range of $Ni_2MnGa$ off-stoichiometries [23]. This phase transition was later obtained in Ni-Mn-Cu-Ga alloys [31, 32], in particular, in $Ni_{50}Mn_{17.5}Cu_{7.5}Ga_{25}$ [33]. More recently, Takhsha *et al*. observed that $Ni_{49.7}Mn_{18.7}Cu_{6.4}Ga_{25.2}$ alloy shows a PM austenite – FM martensite phase transformation around body temperature and reported on the use of FSMA ball-milled particles for hyperthermia, being a novel non-invasive cancer therapy method [25, 26]. Additionally, transition merging within the operational temperature range stands out as a possible strategy to make Ni-Mn-Cu-Ga compounds, and the use of first-order transitions, competitive for thermomagnetic harvesting, an emerging technology to convert waste heat into electricity [34-41].

Since FSMAs exhibiting magneto-structural phase transitions are promising for diverse applications, a comprehensive procedure for characterizing magnetic properties and phase transformations should be established to support experimental work, even in the scenario of direct transitions, i.e., magneto-structural transformations from PM austenite – FM martensite. Here, we synthesize a series of Ni-Mn-Cu-Ga alloys, measure their magnetic properties, and apply a thermodynamic model to quantitatively analyze the data. This allows us to: (i) classify the alloys into three transition types, (ii) extract the virtual Curie temperatures, and (iii) demonstrate that the model captures composition-driven changes in spin-exchange interaction.

## 2. MATERIALS CHARACTERISATION

### 2.1. Experimental

A series of five samples, with nominal compositions $Ni_{50}Mn_{25-x}Cu_xGa_{25}$ ($x$ = 6.25, 6.5, 6.75, 7) and $Ni_{50.5}Mn_{18.5}Cu_{6.5}Ga_{24.5}$, was synthesized by arc melting under an Ar atmosphere. The chemical compositions were chosen to bring the PM–FM phase transition close to the martensitic transformation (MT) temperature range and possibly to obtain a large magnetization jump over the direct transformation from PM austenite to FM martensite. High purity (99.99%) Ni, Mn, Cu and Ga metals were weighed in the appropriate amounts to produce samples of 4 g. To compensate for evaporation losses during melting, 1.5% excess of Mn was added based on prior laboratory experience. Four melting steps were performed, followed by annealing at 1123 K for 72 hours and subsequent water quenching to improve sample homogeneity. The morphological and compositional characteristics of the samples were investigated using a field-emission scanning electron microscope (FE-SEM; Zeiss Auriga Compact by Carl Zeiss) equipped with an energy-dispersive X-ray spectroscopy (EDX) system (Oxford Xplore 30 by Oxford Instruments). The actual stoichiometries were determined by a standardless EDX analysis, using a 20 kV accelerated primary electron beam focused on flat polished sample surfaces. Within the analyzed surface areas, all synthesized samples are compositionally uniform, with elemental contents close to the nominal values (see Table 1). The temperature dependence of the magnetization of the alloys was measured using an extraction magnetometer (Maglab2000 System by Oxford Instruments) and a superconducting quantum interference device (MPMS-XL5 SQUID magnetometer by Quantum Design). Samples were heated prior to the application of the magnetic field. Cooling and heating runs were then measured under an applied magnetic field of 1 T, with a temperature sweep rate of 2 K/min. Magnetic susceptibility as a function of temperature was measured in a thermomagnetic analyzer (TMA) under an AC field of 1 mT. The results of magnetic measurements were subjected to a quantitative theoretical analysis, described in sections 2.2 and 2.3, to characterize the magnetic and phase transformations of the alloys.

| Sample | Nominal composition | EDX composition | e/a |
|---|---|---|---|
| A1 | $Ni_{50}\mathbf{Mn_{18.50}Cu_{6.5}}Ga_{25}$ | $Ni_{49.0}\mathbf{Mn_{19.5}Cu_{6.9}}Ga_{24.6}$ | 7.762 |
| A2 | $Ni_{50}\mathbf{Mn_{18.75}Cu_{6.25}}Ga_{25}$ | $Ni_{49.4}\mathbf{Mn_{19.7}Cu_{6.5}}Ga_{24.4}$ | 7.766 |
| A3 | $Ni_{50}\mathbf{Mn_{18.25}Cu_{6.75}}Ga_{25}$ | $Ni_{49.5}\mathbf{Mn_{19.2}Cu_{7.1}}Ga_{24.2}$ | 7.801 |

| A4 | $Ni_{50}\mathbf{Mn_{18}Cu_7}Ga_{25}$ | $Ni_{50.5}\mathbf{Mn_{18.6}Cu_{7.4}}Ga_{23.5}$ | 7.871 |
|---|---|---|---|
| A5 | $\mathbf{Ni_{50.5}}Mn_{18.5}Cu_{6.5}\mathbf{Ga_{24.5}}$ | $\mathbf{Ni_{50.2}}Mn_{18.8}Cu_{6.9}\mathbf{Ga_{24.1}}$ | 7.818 |

Table 1. The sample code, nominal compositions, EDX compositions and the average number of valence electrons per atom (e/a) calculated on the measured compositions.

### 2.2. Thermodynamic formalism and basic assumption

As shown in the recent publications Refs. [29, 42], the influence of structural phase transition (SPT) on the saturation magnetization of a ferromagnet can be described by taking into account the minimum conditions for the volume density of Gibbs free energy of the magnetic subsystem of the ferromagnet, $G(T,H)$, expressed as:

$$G(T,H) = -\frac{1}{2}J(T)y^2(T,H) - TS(T,H) - y(T,H)M_S H, \tag{1}$$

where $H$ is an external magnetic field, $y(T,H) = M(T,H)/M_S$, $M(T,H)$ represents magnetization values in the external magnetic field, $M_S = M(0,H_S)$, $H_S$ is the saturation field, the coefficient $J(T)$ characterizes the spin-exchange interaction in the ferromagnet undergoing the SPT. This coefficient can be expressed as:

$$J(T) = J_{ex} + \Delta J(T), \tag{2}$$

where the parameters $J_{ex}$ and $\Delta J(T)$ characterise the spin-exchange interaction in the high-temperature phase and the influence of SPT on the intensity of this interaction, respectively.

The function $y(T,H)$ characterizes the rate of ferromagnetic ordering: the value $y(T,0)$ is less than unity at finite temperatures and tends to unity in the low-temperature limit. The entropy of the magnetic subsystem of the alloy is related to the function $y = y(T,H)$ as:

$$S(y) = S_{PM} - \frac{1}{2}nk_B[(1+y)\ln(1+y) + (1-y)\ln(1-y)], \tag{3}$$

where $S_{PM}$ is the entropy of the alloy in the paramagnetic state, $n = N/V$, $N$ and $V$ are the number of magnetic atoms and volume of magnetic solid, respectively, $k_B$ is Boltzmann constant. The minimum conditions for Gibbs free energy and Eqs. (1) – (3) result in the following equation:

$$y = \tanh\left[\frac{\Theta(T)}{T}\left(y + \frac{M_S H}{J_{ex} + \Delta J(T)}\right)\right], \tag{4}$$

where:

$$\Theta(T) = T_C[1 + \Delta J(T) / J_{ex}], \tag{5}$$

with $T_C$ being the Curie temperature of the high-temperature phase (for more details see Ref. [43]). The value $\Theta_C \equiv T_C(1 + j_0 / J_{ex})$ is interpreted as the Curie temperature of the low-temperature phase (for more details see Ref. [29]). Equation (4) enables the computation of the temperature dependence of magnetization in the external magnetic field. The parameter $\Delta J(T)$ is modelled by the function:

$$\Delta J(T) = \frac{1}{2} j_0 \left[ 1 + \tanh\left( \frac{T_{tr} - T}{\Delta_{tr}} \right) \right], \tag{6}$$

where $j_0 = const$, $T_{tr} = T_M$ or $T_{tr} = T_A$ for the forward or reverse MT, respectively, and the constant $\Delta_{tr}$ is about half of the temperature interval of MT. Equation (6) satisfies the inequality $0 < \Delta J(T) < j_0$. This function is considered proportional to the volume fraction of martensite.

### 2.3. Underlying physics

The field-induced change of the absolute value of the magnetization vector **M** is caused by the field-induced increase of the level of ferromagnetic order. If the ferromagnetic ordering appears in the stable martensitic phase or in the stable austenitic phase, the influence of SPT on the spin-exchange interaction is described by the explicit function of temperature $\Delta J(T)$ [29]. A theoretical description of the magnetic properties of FSMA undergoing the ferromagnetic ordering in the mixed austenite-martensite state is an open problem.

For Heusler FSMAs undergoing the martensitic transformation, the high-temperature and low-temperature phases are the cubic phase (austenite) and the low-symmetry phase (martensite), respectively. The value of the parameter $\Delta J(T)$ is proportional to the relative volume change accompanying the MT [30]. For the majority of such alloys the forward MT results in the decrease of the volume of the FSMA and the increase of spin-exchange energy, due to the negative value of the coefficient of proportionality. Therefore, the inequalities $\Delta J(T) > 0$ and $\Theta_C - T_C > 0$ take place [30].

In the martensitic phases of Heusler FSMAs, the crystal lattice has tetragonal, orthorhombic or monoclinic symmetry, but in the first approximation it can be considered as a slightly deformed tetragonal lattice and the martensite state of the alloy can be considered as an ensemble of the spatial domains of the crystal lattice with the fourfold symmetry axes directed along one of the coordinate axes (domains of $x$-variant, $y$-variant and $z$-variant of the martensite phase). Let the magnetic field be applied in the $z$-direction. The application of a magnetic field increases the absolute value of vector **M** of the z-variant but changes both the absolute value and the direction of the magnetization vector **M** of the $x$-variant and $y$-variant of martensite. The direction of the vector **M** can be characterized by the angle $\phi$ between this vector itself and the magnetic field; the approximate relationship $\cos\phi = H / H_S$ can be used for computations, and the saturation field value $H_s$ can be evaluated from magnetization curves $M_z(H)$. The magnetization value of the ensemble of martensite variants is expressed as

$$M_{\text{mart}}(T,H) = \frac{1}{3}M(T,H) + \frac{2}{3}M(T,H)\cos\phi, \quad (7)$$

where:

$$M(T,H) = M_S y(T,H), \quad (8)$$

and $y(T,H)$ is a solution of Eq. (4). Equation (7) is obtained assuming that the statistical weight of every martensite variant in the ensemble of spatial domains is equal to 1/3. The magnetization of a FSMA undergoing the MT is expressed as:

$$\langle M \rangle = \alpha(T)M_{\text{mart}}(T,H) + [1-\alpha(T)]M_{\text{aust}}(T,H), \quad (9)$$

where $\alpha(T) \equiv \Delta J(T) / j_0$ is the volume fraction of the martensite. If the external magnetic field exceeds the saturation value inherent to austenite/martensite, the magnetization of austenite/martensite can be computed from Eq. (8).

The magnetic susceptibility of the alloy can be computed from the following equation:

$$\chi(T,H) = \frac{\partial}{\partial H}\langle M(T,H)\rangle_T . \quad (10)$$

Using the Eqs. (1) – (10), it is possible to consider two mechanisms underlying the magnetization process under the applied magnetic field: first, the ordering of magnetic moments, which are disordered due to the thermal motion of atoms; second, the rotation of the vector sum of atomic magnetic moments toward the magnetic field vector. The first process can be described quantitatively using the fundamental equations of thermodynamics and theory of magnetism (in the given case, Eqs. (1) – (4)). The second process can be described only qualitatively, because of the simplifying assumptions made to derive Eqs. (7) – (10). The second process ends when $H = H_S$ ; therefore, the present theory provides a quantitative description of the magnetization $M(T)$ for $H > H_S$, describes the main features of the magnetic susceptibility $\chi(T)$ measured at $H < H_S$, but is not intended to provide its precise evaluation. For the cubic austenite, the saturation field is of the order of 100 Oe [44]; for the martensite, the saturation field is of the order of 10 kOe.

## 3. RESULTS

The experimental temperature dependences of magnetization, measured in a saturating magnetic field of 10 kOe, are shown in Figs. 1 (a) – (e) in comparison with the theoretical dependences of $\langle M \rangle$, computed from equation $\langle M \rangle = M_S y(T,H)$ and Eq. (4). The low-temperature magnetization value $M_S$, the phase transition temperatures and the characteristic parameters of spin-exchange interaction, which provide the best agreement between theoretical curves and experimental data, are shown in Table 2. $T_C$ and $\Theta_C$ are the Curie temperatures of austenite and martensite, respectively; $T_M$ and $T_A$ are the transformation temperatures of the forward and reverse martensitic transition, respectively.

Figure 1 and Table 2 show that:

i) The alloys A1–A5 can be classified into three types according to the sequence of their magnetic and structural phase transitions. Type I (A1, A2) alloys undergo a magnetic transition from PM austenite to FM austenite upon cooling below the Curie temperature $T_C$, followed by a forward MT to ferromagnetic martensite at lower temperatures $T_M < T_C$ (see Fig. 1(a, b)). On heating, alloy A1 exhibits the reverse sequence (FM martensite → FM austenite → PM austenite), whereas in alloy A2 the reverse martensitic transformation temperature coincides with the Curie temperature ($T_A = T_C$), such that FM martensite transforms directly into PM austenite without an intermediate FM austenite state. Type II (A3, A4) alloys undergo a direct transition from PM austenite to FM martensite on cooling below the martensitic transformation temperature $T_M \geq T_C$ (see Fig. 1(c, d)). During heating, alloys A3 and A4 transform directly from FM martensite to PM austenite. Type III (A5) alloy undergoes a forward martensitic transformation from PM austenite to PM martensite during cooling, followed at lower temperature by a PM–FM magnetic transition in the martensitic phase; the reverse sequence occurs on heating (see Fig. 1(e)).

ii) The martensitic transformation of each alloy results in the abrupt increase of the magnetization value.

iii) The Curie temperatures of austenite and martensite are not noticeable in the magnetization curve if $T_M > T_C$ (see Fig. 1(d, e)); instead, these Curie temperatures can be only determined by fitting theoretical magnetization curves to the experimental data. These are therefore called virtual Curie temperatures. [30]

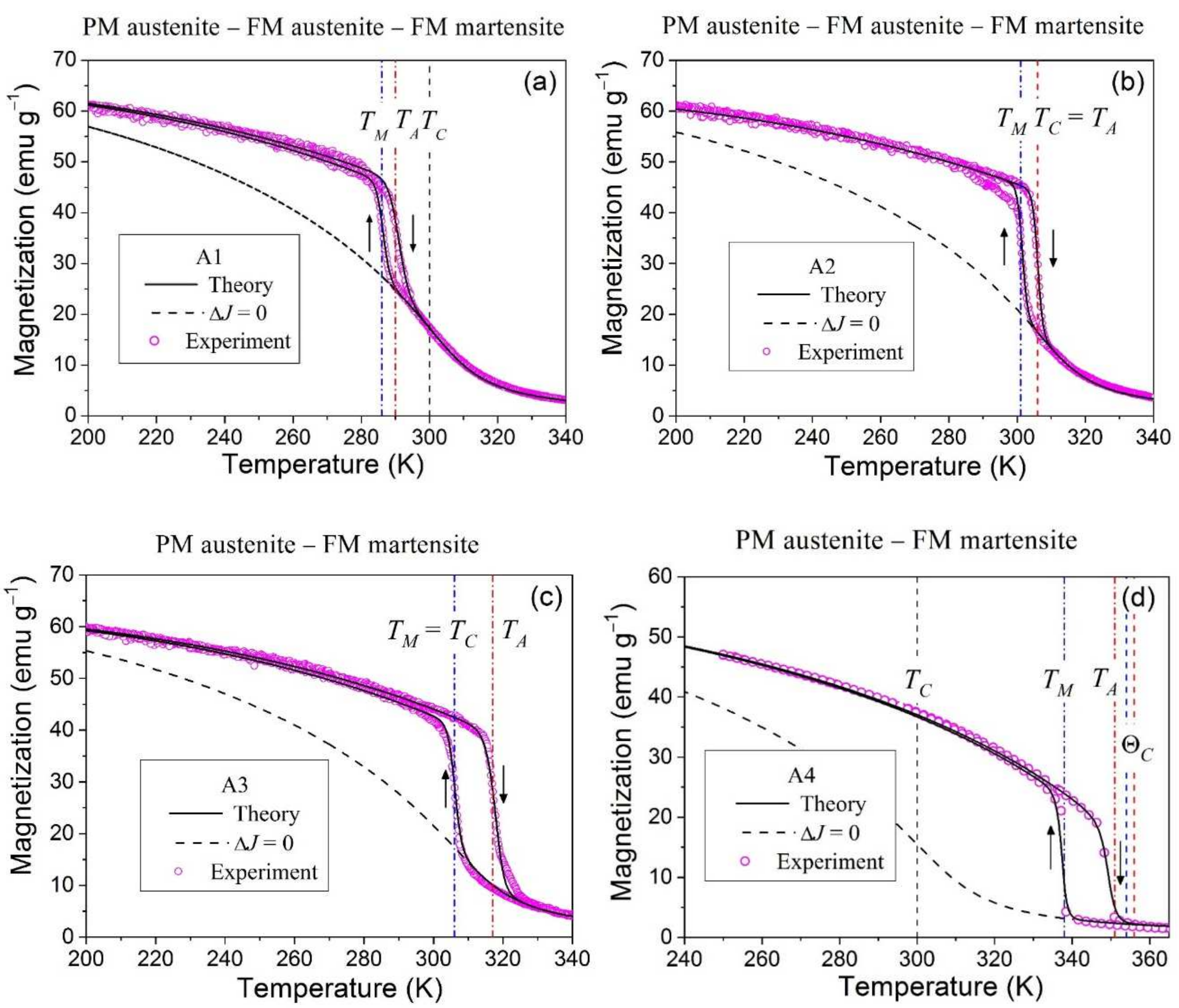

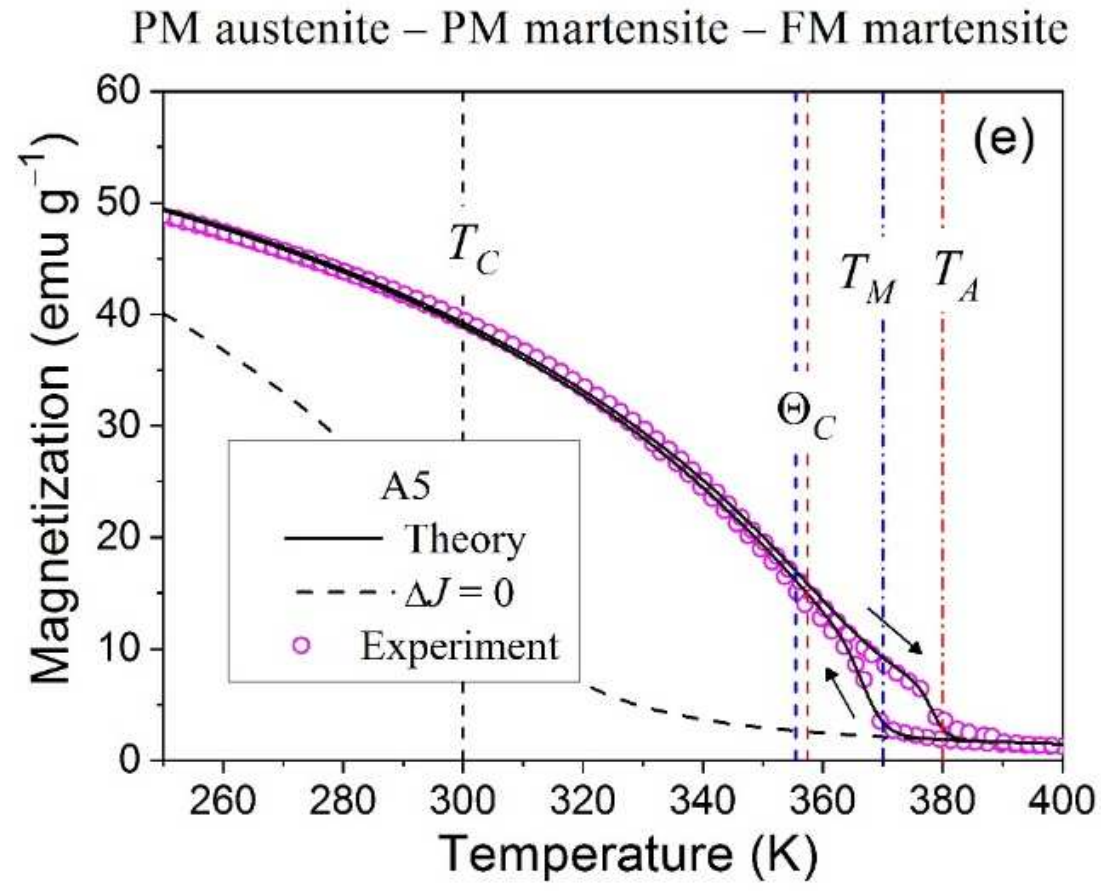

Figure 1. Magnetization curves measured under a saturating magnetic field of 10 kOe of the shape memory alloys undergoing the forward martensitic transformation at: (a), (b) $T_M < T_C < \theta_C$ ; (c),(d) $T_C \leq T_M < \theta_C$; (e) $T_M > \theta_C > T_C$. The dashed lines show magnetization functions computed for ferromagnets without martensitic transformations, for comparison.

| | A1 | A2 | A3 | A4 | A5 | |
|---|---|---|---|---|---|---|
| $T_M$ (K) | 286 | 301 | 306 | 338 | 370 | Cooling |
| $T_A$ (K) | 290 | 306 | 317 | 351 | 380 | Heating |
| $T_C$ (K) | 300 | 306 | 306 | 300 | 300 | |
| $\Theta_C$ (K) | 351 | 373 | 358 | 354 | 356 | Cooling |
| | 357 | 373 | 367 | 356 | 357 | Heating |
| $M_S$ (emu/g) | 66 | 64 | 63.3 | 56.7 | 59.3 | |
| $J_{ex}/M_S$ (Oe) | $1.6\cdot10^6$ | $1.7\cdot10^6$ | $1.4\cdot10^6$ | $1.4\cdot10^6$ | $1.2\cdot10^6$ | |
| $j_0/J_{ex}$ | 0.17 | 0.22 | 0.18 | 0.175 | 0.185 | Cooling |
| | 0.19 | 0.22 | 0.2 | 0.185 | 0.19 | Heating |
| $\Delta_{tr}$ (K) | 2 | 1.5 | 2 | 1 | 4 | Cooling |
| | 3 | 1.5 | 3 | 3 | 2.5 | Heating |

Table 2. Phase transition temperatures and characteristic parameters resulting in the theoretical magnetization curves shown in Fig. 1. $T_M$: forward martensitic transformation temperature; $T_A$: reverse martensitic transformation temperature; $T_C$: Curie temperature of the austenitic phase; $\theta_C$: Curie temperature of the martensitic phase; $M_S$: saturation magnetization; $J_{ex}/M_S$: ratio between the exchange parameter and the saturation magnetization; $j_0/J_{ex}$: ratio between the exchange constant and the exchange parameter value; $\Delta_{tr}$: constant value corresponding to about half of the temperature interval of the martensitic transition.

The derivatives $\partial\langle M\rangle/\partial T$ are shown in Figs. 2 (a) – (e) together with the magnetic susceptibilities computed from Eq. (10). Figure 2 and Table 2 enable the main conclusions to be drawn from the theoretical analysis of experimental temperature dependences of magnetization measured for the alloys with adjusted MT temperature:

I) If a FSMA undergoes a martensitic transformation below the Curie temperature of the austenite, the latter corresponds to the second-order PM – FM phase transition. The PM – FM phase transition temperature can be determined from the maximum of magnetic susceptibility, while the MT temperature can be found from the minimum of the temperature derivative of magnetization (see Fig. 2 (a)). The Curie temperature of the martensite does not correspond to the maximum of magnetic susceptibility or to the minimum of $\partial\langle M\rangle/\partial T$. The Curie temperature of the austenite is the real Curie temperature of FSMA, while the Curie temperature of the martensite is referred to as the virtual Curie temperature [30].

II) If MT temperatures are higher than the Curie temperature of austenite and lower than the Curie temperature of martensite, the direct magneto-structural phase transitions PM austenite – FM martensite (on cooling) and FM martensite – PM austenite (on heating) take place. The temperatures of these phase transitions can be estimated from the temperature dependences of magnetic susceptibility and magnetization (see Fig. 2 (d)). According to its definition, the Curie temperature corresponds to a temperature at which the PM austenite → FM martensite transition starts upon cooling, or at which the FM martensite → PM austenite transition finishes upon heating. The Curie temperatures of austenite and martensite do not correspond to the maximum of the magnetic susceptibility or the minimum of $\partial\langle M\rangle/\partial T$.

III) If a FSMA undergoes martensitic transformation above the Curie temperature of martensite, this Curie temperature is the temperature of the PM – FM phase transition. The PM – FM phase transition temperature can be determined from the maximum of magnetic susceptibility. The MT temperature can be roughly estimated from the minimum of the temperature derivative of magnetization; however, the difference between this estimate and the MT temperature corresponding to the best fit of theoretical curves to experimental data is about 3 K for the forward MT and 2 K for the reverse MT (see Fig. 2(e)). The Curie temperature of austenite does not correspond to the maximum of magnetic susceptibility or to the minimum of $\partial\langle M\rangle/\partial T$. The Curie temperature of martensite is the real Curie temperature of FSMA, while the Curie temperature of austenite is referred to as the virtual Curie temperature [30].

It should be emphasized, first, that the noticeable difference between the MT temperatures providing the best fit of theoretical $M(T)$ curves to experimental points and the temperatures corresponding to the minima of derivatives $\partial\langle M\rangle/\partial T$ is caused by the fact that the temperature anomaly observed at $M(T)$ curve is less pronounced in A4 than in the other alloys, and therefore, the fitting of theoretical curve to experimental points does not provide high accuracy. Second, it should be remembered that the martensitic transformations of the alloys A1 – A5 intensify the spin-exchange interaction [29]. In this case the inequality $\Theta_C > T_C$ takes place. The opposite case needs special consideration.

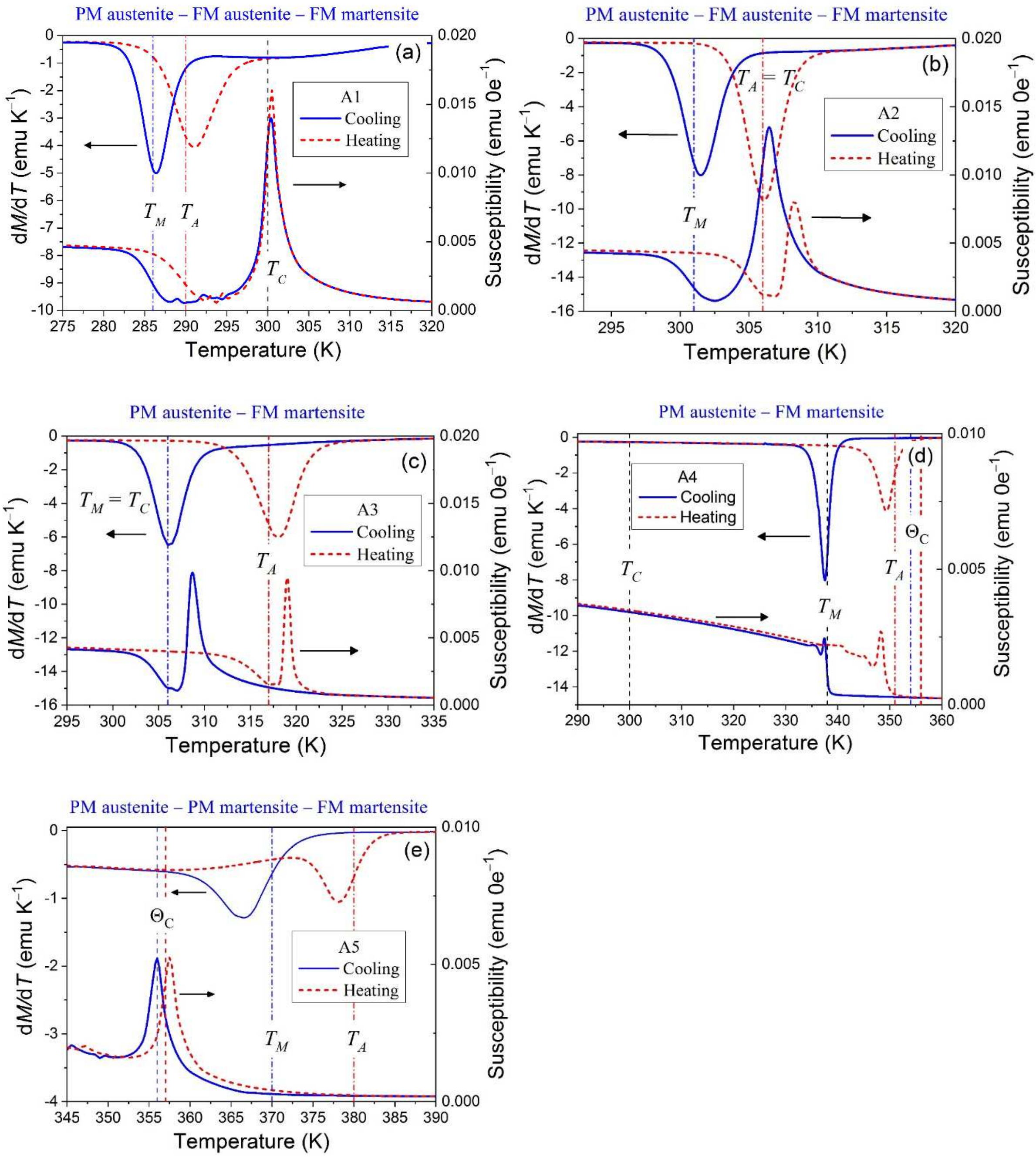

Figure 2. Magnetic susceptibility computed for the field value 100 Oe and the temperature derivative of magnetization for a magnetic field of 10 kOe of the shape memory alloys undergoing the forward martensitic transformation at: (a), (b)$T_M < T_C$; (c), (d) $T_M \geq T_C$; and (e) $T_M > \theta_C$.

Figure 3 shows the experimental temperature dependences of magnetic susceptibility, $\chi(T)$, obtained for the alloys A1 – A5, belonging to types I), II), III) respectively. The insets in the figure illustrate the resemblance between the experimental $\chi(T)$ loops and the $\chi(T)$ loops computed using the model equations (7-10). Obvious discrepancies in the shapes of experimental and theoretical loops are caused, it seems, by the fact that the equation (7) is applicable to the martensite structure arising on cooling of Ni-Mn-Ga-Cu single crystals but disregards the peculiarities of the magnetic properties of the polycrystalline samples. Indeed, the maximum values of magnetic susceptibility of A1 and A5 alloys are observed well below the Curie temperatures of austenite and martensite. The temperature hysteresis of the magnetic susceptibility of A5 is approximately equal to the difference of the $\Theta_C$ values resulting from the computed temperature dependence of magnetization (see Fig. 2 (e)), but the temperature hysteresis of the magnetic susceptibility of A1 is qualitatively different from that predicted by the theory. The theoretical $\chi(T)$ curves corresponding to the cooling and heating runs computed in the magnetic saturation field do not show significant hysteresis in the neighborhood of the Curie point (see Fig. 2 (a)), while the $\chi(T)$ curves measured in the much lower field practically coincide when the magnetic susceptibility decreases, but show a significant hysteresis otherwise (see Fig. 3 (a)). It should also be noted that the temperature dependences of the magnetic susceptibility for alloys A2 and A3 (Fig. 3(b, c)) differ qualitatively from those predicted by theory. These facts point to the dependence of magnetic susceptibility on the microstructure of martensite, because the microstructure of Ni-Mn-Ga-Cu martensite arising in the magnetic saturation field can be noticeably different from that appearing in the low field.

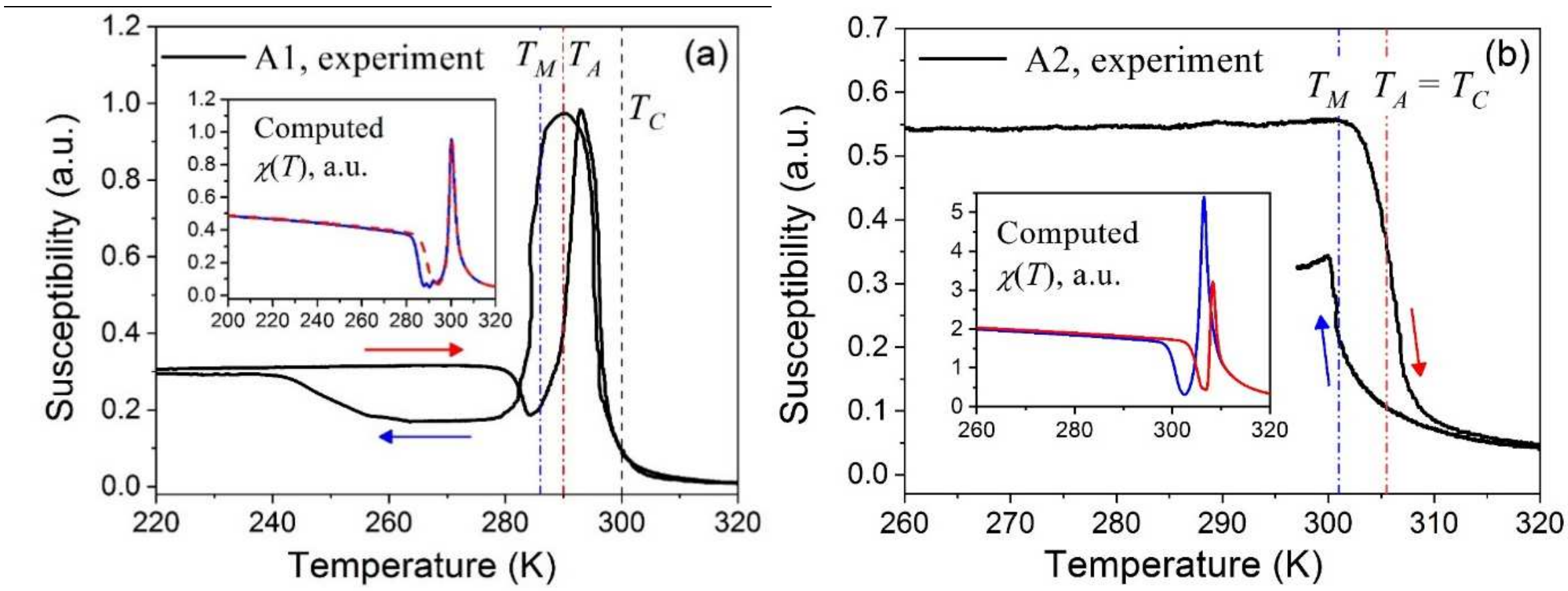

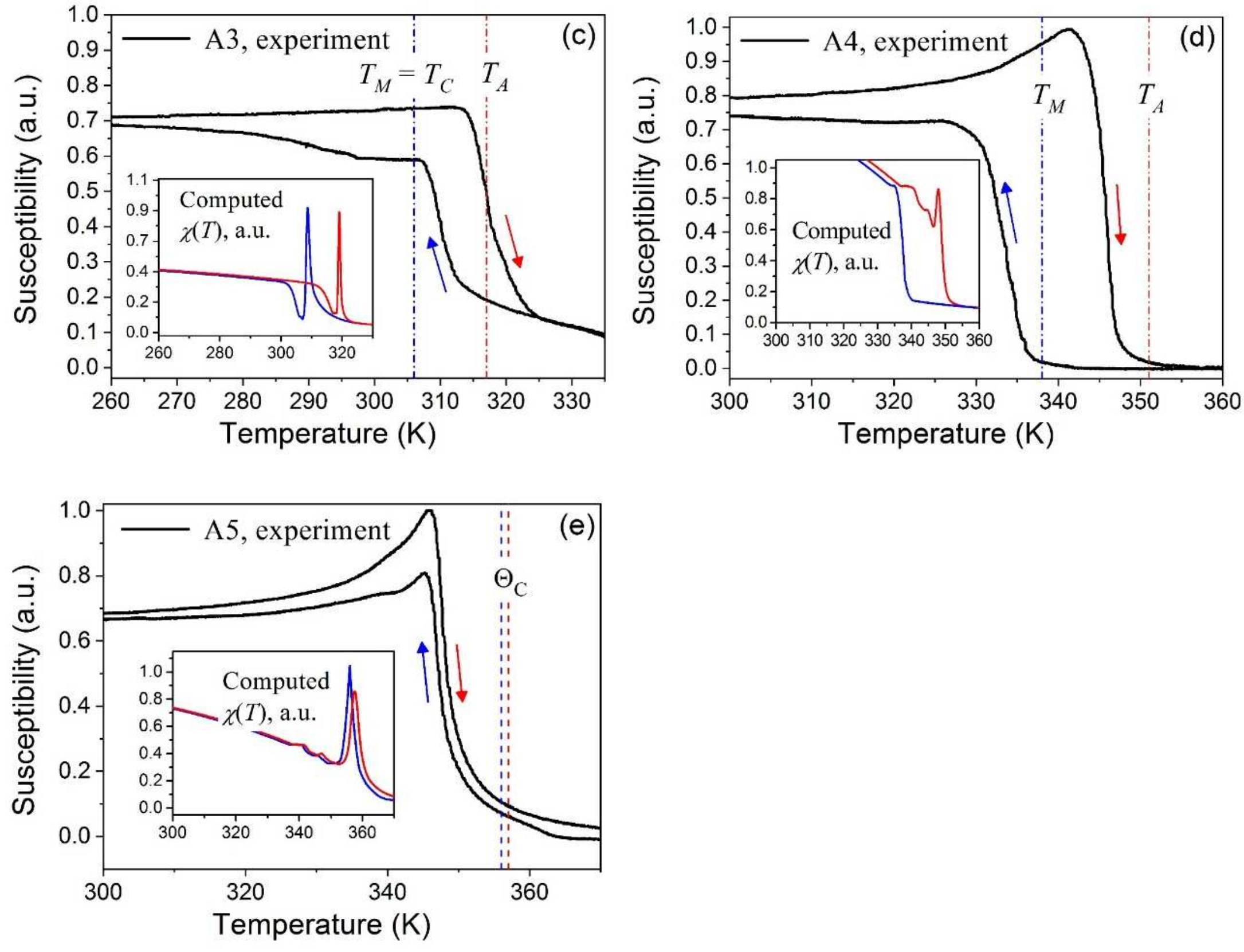

Figure 3. Experimental temperature dependences of magnetic susceptibility of the alloys: (a), (b) type I; (c), (d) type II; and (e) type III. Insets: theoretical temperature dependences computed with a 0.5 K step, close in value to experimental temperature step.

## 4. DISCUSSION

Copper was purposefully added to $Ni_{50}Mn_{25}Ga_{25}$ to bring the MT close to the Curie point of these alloys over a temperature window spanning from room temperature to 400 K. According to the literature, Cu substitution of Mn [31] or Ga [32] increases the MT temperature and decreases the Curie temperature of the austenitic phase ($T_C$). Consequently, there exists a compositional range where the alloy undergoes a first-order direct magneto-structural transition from a FM martensite to a PM austenite upon heating, and vice versa upon cooling. For higher Cu contents, the martensitic phase remains stable up to almost 500 K, and its Curie temperature ($\Theta_C$) decreases with increasing Cu [31]. The occurrence of a direct transition in Ni-Mn-Ga can be explained by introducing virtual Curie temperatures (austenitic Curie temperature = $T_C$ and martensitic Curie temperature = $\Theta_C$) [23, 30]: in this framework, the necessary condition is $T_C < T_M < \Theta_C$. In this case the Curie temperatures are called virtual as they cannot be directly detected via experimental measurements. According to this approach, the distance between $T_C$ and $\Theta_C$ becomes an indicator of the width of the chemical composition range where the direct FM martensite to PM austenite transition occurs. A direct transition is observed for samples A3, and A4 in the present set (see Table 2). Sample A2 represents a limiting case, because during heating the reverse MT temperature lies in the range $T_C \leq T_A < \Theta_C$, whereas on cooling it first undergoes a PM-FM transition, and only afterwards does the MT occur, when austenite is already ferromagnetic, i.e. $T_M < T_C$.

Additionally, the decreasing saturation magnetization ($M_s$) from A1 to A4 can be attributed to the Mn depletion trend detected by EDX measurements. In Ni-Mn-Ga-based alloys the Mn atoms carry the highest magnetic moment (~4 μB) [45]: Cu substitution on the Mn site [46] reduces the number of magnetically active Mn atoms contributing to the net magnetization.

Considering the variation of the exchange parameter over the structural transitions (denoted as $\Delta J(T)$) enables the MT temperatures to be estimated with excellent agreement to the experimental data, with a maximum difference between fitted and experimental data of 3 K (see Table 3), and the reproduction of magnetization curve features with high accuracy. Experimental values of MT temperatures were determined as the minimum of the derivatives of measured magnetization curves with respect to temperature. The present theoretical approach also enables the determination of the virtual Curie temperatures, which cannot be derived from experimental data in case of direct transitions [29].

The variation of the exchange parameter $\Delta J(T)$ captures the reorganization of the electronic states that accompanies the lattice transformation during the MT, which is fostered by a complex electron-phonon coupling mechanism. Within the commonly accepted framework, when Fermi surface nesting occurs, the structural deformation associated with the martensitic transition becomes energetically favorable for the system, and a strong phonon-electron coupling occurs at the wave vectors corresponding to the nesting direction. Such directional restorative forces shear the lattice and induce a reorganization of the electronic structure, leading to a partial depletion of the density of states (DOS) near the Fermi level as in a band Jahn-Teller effect like mechanism. In Ni-Mn-Ga based Heusler alloys, the electronic states near the Fermi level are largely dominated by Ni 3d states hybridized with Mn 3d orbitals, making nickel a key contributor to the electronic instability associated with the martensitic transformation [47-53]. Based on previous investigations, Cu-doping of $Ni_{50}Mn_{25}Ga_{25}$ acts indirectly by modifying the local chemical environment of Ni and the interatomic covalent hybridization strength with Mn, Cu and Ga [54, 55]. Overall $Ni_{50}Mn_{25-}$ ${}_xCu_xGa_{25}$ and $Ni_{50}Mn_{25}Cu_xGa_{25-x}$ are more delocalized than $Ni_{50}Mn_{25}Ga_{25}$. Consequently, the lattice is softer, and MT temperatures increase with respect to the undoped $Ni_{50}Mn_{25}Ga_{25}$ [54]. This increase is described phenomenologically, within the present theory, by introducing the temperature-dependent parameter $\Delta J(T)$. The modification of the electronic structure in response to the stoichiometry tweaking may be rationalized in terms of variation in the average number of electrons per atom (e/a), which in metamagnetic Heusler compounds have been reported to correlate approximately linearly with transition temperatures [9] over certain compositional ranges. In the present set, the error associated with the measured EDX compositions hinders the evaluation of transition temperature trends as a function of e/a, and Ni atomic percentage. Measurements were performed following a standardless approach, normally leading to a relative error of ~ 5% for major elements (>10 wt% or higher), and 7-10%, for minor constituents like Cu (6 - 7 wt%) [56]. Moreover, further experimental work would be needed to investigate other factors, like the degree of chemical order and the way the different atomic species occupy the available lattice sites. Consequently, we restrict our discussion to qualitative compositional regimes: Cu-poor (before the direct transition sets in: type I), Cu-rich (compositions showing direct transition: type II), Ni-rich (type III), rather than fine e/a scaling laws. Within this limitation, the proposed thermodynamic model detects composition-dependent electronic structure modifications over the present A1-A5 series through the spin-exchange interaction and the magnitude of $\Delta J(T)$ in the different samples, which are consistent with composition-driven modification of transitions suggested in previous studies [31, 32, 54, 55]. In this sense, the model provides a complementary approach, which is promising for the study of the relation between the alloy composition and its magnetic properties when encountering experimental limitations caused by, e.g., EDX sensitivity, which hinders a precise evaluation of small stoichiometric variations. Results obtained for sample A5 ($T_C \leq \Theta_C < T_M$) suggest that excess Ni plays a crucial role in the stabilization of the martensitic phase up to room temperature, in compliance with previous results reported on off-stoichiometric Ni-Mn-Ga systems [45, 57, 58].

The proposed thermodynamic analysis also stands out as a suitable approach for determining virtual Curie temperatures in the systems where MT is close to the Curie point, which is an open problem. Virtual Curie temperatures were determined by fitting theoretical magnetization curves, computed based on the analytical model, to the experimental data. $T_C$ values do not show a monotonic trend over the compositional range but are instead rather constant. A possible explanation is the effect of volume magnetostriction, which is inherent to these alloys [30]. When the austenitic Curie temperature lies close to the martensitic transformation temperature, the volume magnetostriction associated with magnetic ordering contributes to lattice strain. This magnetoelastic coupling can modify the stability of the structural phases and changes the compositional interval where magneto-structural transitions occur. In this case the interval of existence of magneto-structural phase transition depends on the magnitude of the volume magnetostriction. This is reflected in the difference between Curie temperatures of austenite and martensite and justifies the existence of magneto-structural transitions over large stoichiometric ranges [30, 45, 59]. Such a coupling may reduce the sensitivity of the austenitic Curie temperature to small compositional variations: the spontaneous magnetization appears at MT temperature, which can be rather far from Curie temperatures. Therefore, the volume magnetostriction couples the magnetic and structural degrees of freedom and results in the difference between the Curie temperatures of the austenite and the martensite [30, 33].

$\Theta_C$ values show larger variations but not a clear trend as well. The Curie temperature of martensite is obtained directly from Eq. (5) and shows different values between cooling and heating (see Table 2). These are predetermined by the parameter $j_0/J_{ex}$. The slight variations in $j_0/J_{ex}$ between cooling and heating runs (Table 2) reflect the fact that the reverse MT takes place in the presence of internal strains/stresses caused by the martensite microstructure. Therefore, the $\Theta_C$ temperature computed for the heating run corresponds to the phase transition from mechanically stressed martensite to austenite. This demonstrates the sensitivity of the fitting procedure and the importance of using consistent protocols for heating/cooling measurements.

Computed susceptibility curves are rather different from experimental plots. This is because the model assumes, for simplicity, equal statistical weights of 1/3 for the three martensite variants, which is an approximation that neglects microstructure in polycrystalline samples. However, the model can successfully predict the drastic influence of small variations of the chemical composition on the temperature dependence of the magnetic susceptibility of a given alloy, also highlighted by experimental results. For example, referring to the three transition kinds defined earlier, the magnetic susceptibility of $Ni_{50}Mn_{18.50}Cu_{6.5}Ga_{25}$ (A1, type I) alloy, undergoing the PM – FM phase transition in the austenitic state, reaches its maximum at the Curie temperature of the austenite and reaches a minimum below this temperature; the direct magneto-structural phase transition in $Ni_{50}Mn_{18}Cu_{7}Ga_{25}$ (A4, type II) alloy results in the step-wise decrease of magnetic susceptibility; the magnetic susceptibility of $Ni_{50.5}Mn_{18.5}Cu_{6.5}Ga_{24.5}$ (A5, type III) alloy, undergoing the PM – FM phase transition in the martensitic state, reaches its maximum at the Curie temperature of martensite but exhibits no minimum below this temperature.

Another important aspect to consider for first-order transitions is hysteresis. Much research is focused on understanding how to minimize it to improve the applicability of materials for cyclic applications. Reduced hysteresis results from an improved compatibility between the austenitic and martensitic crystalline phases, which reduces the stress generated over the phase transition, thus improving the fatigue life [60, 61]. Moreover, reducing hysteresis would increase the reversibility of the structural transition, consequently improving the efficiency of cyclic caloric applications [22]. In this context, Mendonça *et al.* suggested that in Heusler alloys, where Mn substitution leads to direct transitions, the hysteresis values first decrease up to the critical substituting-element concentration, i.e. just before direct transitions occur, then increase at higher concentrations where direct transitions are present. The trend is valleylike and generally reaches a minimum where both phases, austenite and martensite, are either paramagnetic or ferromagnetic [60]. Hysteresis values over the present sample series follow the same pattern (see Table 3). Hysteresis is small for samples A1 and A2, showing the PM-FM phase transition in the austenite state, while it increases consistently in samples A3 and A4, which undergo direct transitions between a PM austenite and a FM martensite.

| | sample | A1 | A2 | A3 | A4 | A5 | |
|---|---|---|---|---|---|---|---|
| *theoretical* | $T_M$ (K) | 286 | 301 | 306 | 338 | 370 | Cooling |
| | $T_A$ (K) | 290 | 306 | 317 | 351 | 380 | Heating |
| | Hys (K) | 4 | 5 | 11 | 13 | 10 | |
| *experimental* | $T_M$ (K) | 286 | 302 | 306 | 338 | 367 | Cooling |
| | $T_A$ (K) | 291 | 306 | 317 | 349 | 377 | Heating |
| | Hys (K) | 5 | 4 | 11 | 11 | 10 | |

Table 3. Comparison between MT temperature and hysteresis values derived from theoretical calculations and experimental magnetization curves. $T_M$ denotes the forward martensitic transformation temperature; $T_A$ denotes the reverse martensitic transformation; Hys denotes the hysteresis value calculated from the theoretical and experimental MT temperatures, respectively reported in rows 1 and 2.

## 5. SUMMARY AND CONCLUSION

Ferromagnetic Cu-doped $Ni_{50}Mn_{25}Ga_{25}$ samples were synthesized and investigated as they show direct transitions from ferromagnetic martensite to paramagnetic austenite at suitable temperatures for multifunctional applications. The sample set ($Ni_{50}Mn_{25-x}Cu_xGa_{25}$ with $x = 6.25$, 6.5, 6.75, 7) and $Ni_{50.5}Mn_{18.5}Cu_{6.5}Ga_{24.5}$ allowed for the thermodynamic analysis of three kinds of transitions: (type I) the paramagnetic-ferromagnetic phase transition in the austenite state; (type

II) the direct magneto-structural phase transition from paramagnetic austenite to ferromagnetic martensite; (type III) the paramagnetic-ferromagnetic phase transition in the martensite state. The thermodynamic model consistently captures the interplay between structural and magnetic transitions and provides a reliable framework for determining real and virtual Curie temperatures in Heusler alloys by quantitatively reproducing the temperature dependence of magnetization and qualitatively reproducing the temperature dependence of susceptibility, even in the complex scenario of direct magneto-structural transitions. Indeed, when the MT approaches the Curie point of a ferromagnet, its magnetic and caloric properties are strongly affected, and show a sharp dependence on the temperature difference between the martensitic and Curie temperatures [29].

Overall, the obtained results demonstrate that small compositional variations strongly affect magnetic susceptibility and transition sequences. The quantitative theoretical analysis of the experimental temperature dependences of magnetization, $M(T)$, shows that this fact is caused by the impact of martensitic transformation on the spin-exchange energy, which moreover results in a large difference ($\geq$ 50 K) between the Curie temperatures computed for the austenite and martensite. This large difference is mainly caused by a shift in the Curie temperature of the martensite, while the variation in the austenitic Curie temperatures due to the different chemical compositions of the alloys appeared to be one order of magnitude smaller (see Table 2).
Most importantly, the thermodynamic analysis highlights the necessity of using different approaches to derive characteristic temperature values around magneto-structural phase transitions. The characteristic temperatures of the direct magneto-structural phase transition are close to the minima of $\partial\langle M\rangle/\partial T$ function. In contrast, the Curie points of PM – FM phase transitions occurring in austenitic and martensitic states are not the peculiar points of $M(T)$ and $\partial\langle M\rangle/\partial T$ graphs, but they are close to the maximum points of magnetic susceptibility, which can be determined experimentally (using the standard equipment) or theoretically (from the experimental dependences $M(T)$).

The good agreement between computed and experimental temperature values validates the solidity of the proposed theoretical model, which stands as a robust framework for interpreting complex magnetization data, and extracting physically meaningful transition temperatures that cannot be obtained directly from experiments alone. It suggests the importance of the influence of SPT on the spin-exchange parameter. This finding is relevant for a proper investigation of multiferroic systems, specifically, to determine real and virtual Curie temperatures from standard magnetization data including when they cannot be measured directly. This is a significant contribution to the characterization of materials with direct transitions. The methodology developed here is readily applicable to other ferromagnetic Heusler systems and a transferable framework for the analysis of other multiferroic ferromagnetic materials, supporting the rational design of alloys for multifunctional applications, such as energy conversion, actuation, and biomedical technologies.

**Acknowledgements**

V.A.L. acknowledges a financial support from the National Academy of Science of Ukraine (Project 0124U000392).

A.K. acknowledges a financial support from the Austrian Science Fund (FWF) [10.55776/RIC5537124] and the European Innovation Council, funded by the European Union, through project CoCoMag (Grant No. 101099736).

IMEM-CNR and the University of Parma and T.S. acknowledge the Heat4Energy project. This project has received funding from the EU under grant agreement No 101119852. Views and opinions expressed are however those of the author(s) only and do not necessarily reflect those of the European Union or European Innovation Council and SMEs Executive Agency (EISMEA) or European Research Executive Agency (REA). Neither the European Union nor the granting authority can be held responsible for them.